\documentclass[pra]{revtex4-1}

\usepackage{amssymb,amsmath,xcolor,graphicx,xspace,colortbl,rotating} %
\usepackage[raggedrightboxes]{ragged2e} 
\usepackage{amsfonts}  
\usepackage{amsmath}  
\usepackage{amssymb}  
\usepackage{boxedminipage}  
\usepackage{float}  
\usepackage{graphics}  
\usepackage{graphicx}  
\usepackage{ragged2e}  
\usepackage{tabulary}  
\usepackage{wrapfig}  
\usepackage{xcolor}  
\graphicspath{{tunablemetasurface_graphics/}{tunablemetasurface_tcache/}{tunablemetasurface_gcache/}}
\DeclareGraphicsExtensions{.pdf,.eps,.ps,.png,.jpg,.jpeg}
\providecommand{\U}[1]{\protect\rule{.1in}{.1in}}

\begin{document}

\title{Tunable subnanometer gap plasmonic metasurfaces
}
\author{Dennis Doyle}
\affiliation{NREIP Summer Student from University of Pittsburgh, Naval Research Laboratory,~4555 Overlook Ave. Washington, D.C. 20375, U.S.A.}
\author{Christos Argyropoulos}
\affiliation{ONR Summer Faculty Fellow from the University of Nebraska-Lincoln, Naval Research Laboratory,~4555 Overlook Ave. Washington, D.C. 20375, U.S.A.}
\author{Rafaela Nita}
\affiliation{NRC postdoctoral fellow, Naval Research Laboratory,~4555 Overlook Ave. Washington, D.C. 20375, U.S.A.}

\author{Joeseph B. Herzog}
\affiliation{ONR Summer Faculty Fellow from the University of Arkansas, Naval Research Laboratory,~4555 Overlook Ave. Washington, D.C. 20375, U.S.A.}
\author{Nicholas Charipar, Scott A. Trammell, Jawad Naciri, Alberto Piqu{\'e} and Jake Fontana}

\email{jake.fontana@nrl.navy.mil}

\affiliation{Naval Research Laboratory,~4555 Overlook Ave. Washington, D.C. 20375, U.S.A.}

\begin{abstract}The index of refraction governs the flow of light through materials.  At visible and near infrared wavelengths the real part of the refractive index is limited to less than $3$ for naturally occurring transparent materials, fundamentally restricting applications.  Here, we carried out experiments to study the upper limit of the effective refractive index of self-assembled metasurfaces at visible and near-infrared wavelengths.  The centimeter-scale metasurfaces were made of a hexagonally close packed (hcp) monolayer of gold nanospheres coated with tunable alkanethiol ligand shells, controlling the interparticle gap from $2.8$ to $0.45$ $nm$.  In contrast to isolated dimer studies, the macro-scale areas allow for billions of gaps to be simultaneously probed and the hcp symmetry leads to large wavelength shifts in the resonance mode, enabling subnanometer length scale mechanisms to be reproducibly measured in the far-field.  We demonstrate for subnanometer gaps, that the optical response of the metasurfaces agrees well with a classical (local) model, with minor nonlocal effects and no clear evidence of ligand-mediated charge transfer at optical frequencies.  We determine the effective real part of the refractive index for the metasurfaces has a minimum of $1.0$ for green-yellow colors, then quickly reaches a maximum of $5.0$ in the reds and remains larger than $3.5$ far into the near infrared.  We further show changing the terminal group and conjugation of the ligands in the metasurfaces has little effect on the optical properties.  These results establish a pragmatic upper bound on the confinement of visible and near infrared light, potentially leading to unique dispersion engineered coatings.  
\end{abstract}

\volumeyear{year}
\volumenumber{number}
\issuenumber{number}
\eid{identifier}
\date{date}
\received[Received text]{date}
\revised[Revised text]{date}
\accepted[Accepted text]{date}
\published[Published text]{date}
\startpage{101}
\endpage{102}
\maketitle
\bigskip Recent advances in the understanding of the interactions between electromagnetic fields and matter has led to the development of many exciting optical properties from plasmonic nanostructures. These structures separated by nanometer size gaps can confine electromagnetic fields into spatial regions well below the diffraction limit,\cite{RN79} resulting in significant enhancement of the fields,\cite{RN450} potentially leading to novel plasmonic sensors,\cite{RN515,RN591,RN1478} lenses,\cite{RN1441,RN1465} nanolasers,\cite{RN1709,RN1421,RN1006} photovoltaics,\cite{RN1708} and nanoscale antennas.\cite{RN1100,RN1710}

To maximize the confinement of the electric field and ensuing optical properties such as the effective index of refraction,\cite{RN1703,RN1719} classical theory predicts as the gap size between two plasmonic nanospheres approaches zero the local field enhancement can become arbitrarily large, which is not physically possible, therefore breakdown of the field must occur.\cite{RN1713}  The onset for the breakdown of electric field (nonlocal response) in the gap between the nanospheres in vacuum has been shown to experimentally occur at length scales comparable to or less than the atomic lattice constant of gold ($0.41$ nm).\cite{RN1014,RN844} Other experimental work reported if organic ligand spacers are placed in the gap, that nonlocal\cite{RN1415} and full breakdown, allowing charge transfer for gaps as large as $1$ nm, can be observed.\cite{RN1629,RN1628}    Yet, open questions remain as to the mechanisms that would facilitate charge transfer through the ligands at optical frequencies and resulting charge-transfer plasmon modes.\cite{RN1627}

To investigate these subnanometer gap mechanisms, we created large-area metasurfaces using directed self-assembly.  The metasurfaces are comprised of a hcp monolayer of gold nanospheres coated with tunable alkanethiol ligand shells, with interparticle gaps approaching atomic lattice length scales.  We experimentally show that the optical response of the metasurfaces agrees well with a classical model, with only small nonlocal effects.  We demonstrate the effective real part of the refractive index is as small as $1.0$ at $555$ $nm$ and as large as $5.0$ at $743$ $nm$. Finally, we find changes to the optical properties of the metasurfaces are negligible if the terminal group or conjugation of the ligands is varied.

\textbf{RESULTS AND DISCUSSION}

The metasurface geometry (Figure 1a) consists of a monolayer of hcp gold nanospheres coated with ligand shells of tunable length on a glass substrate.\cite{RN590,RN1711} The interparticle gap is controlled by varying the chain length of the alkane thiol ligands.  A mixture of equal length alkane chains with mono- and di-thiols was used.  The monothiol ligands bond to the surface of the gold nanospheres leaving the alkane chain exposed, providing a repulsive (entropic) force aiding to prevent aggregation while dithiols provide an attractive force covalently bonding the nanospheres together.   The chain lengths were varied from C2 to C14, where the $C -$number indicates the number of carbon atoms in the alkane chains.

\begin{figure}\centering 
\setlength\fboxrule{0.01in}\setlength\fboxsep{0.1in}\fcolorbox[HTML]{FFFFFF}{FFFFFF}{\includegraphics[ width=6.5in, height=1.5782196969696969in,]{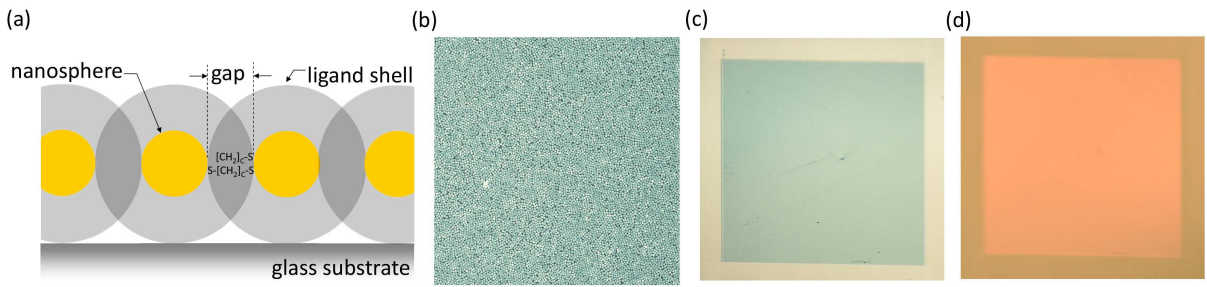}
}
\caption{ (a) Schematic of the metasurface.  (b) False-colored TEM image of a metasurface (C3).  The average nanosphere diameter, $13 \pm 0.7$ $nm$, can be used to scale the TEM image. Transmission (c) and reflection (d) true-color photos of a metasurface (C3).  The metasurfaces were trimmed into $7$ by $7$ $mm$ squares using a UV laser ablation system.
}\label{1}\end{figure}

A representative false-colored TEM image of a metasurface is shown in Figure 1b, validating the metasurfaces are monolayers over macro-length scales while maintaining nanometer sized gaps.  Photographs of a $0.5$ $cm^{2}$ metasurface, backlit with white light, is shown in transmission (Figure 1c) and reflection (Figure 1d).  The gold colored reflection is a consequence of the dense packing of gold nanospheres and the spatially uniform blue color indicates the nanoparticles are not aggregated (touching).

To quantify the interparticle gap for each metasurface, transmission electron microscopy (TEM) grids were prepared by gently scooping the metasurfaces with the grids from the air-fluid interface as they were assembling.  From the TEM imagery, $1 ,000$ gaps were measured for each metasurface.  The nanosphere diameter from all measurements was determined to be $13 \pm 0.7nm$. The distribution of gap sizes is shown in Table 1 and the histograms in Figure 2.  The gaps ranged from $2.8$ $nm$ for C14 down to $0.45$ $nm$ for C2.  The metasurfaces hexgonally close packed at nano- and micro-meter length scales, though the packing was amorphous at larger size scales. The full-width-at-half-maximum ($FWHM$) values from the gap histograms in Figure 2 shows the gap uniformity decreased as the ligand length increased.

In addition to tuning the interparticle gaps using different ligand lengths, the introduction of the dithiol ligands is a significant and key advancement over previous reports,\cite{RN1711} improving the uniformity and area of the metasurfaces (Figure 1) and highlighting the need for both attractive and repulsive forces during the self-assembly process.  Moreover, the dithiols also mechanically strengthened the metasurfaces sufficiently to enable transfer to TEM grids, thereby facilitating direct measurements of the interparticle gaps (Figure 2), which was only inferred for a single gap previously.\cite{RN1711}
\begin{figure}\centering 
\setlength\fboxrule{0.01in}\setlength\fboxsep{0.1in}\fcolorbox[HTML]{FFFFFF}{FFFFFF}{\includegraphics[ width=6.5in, height=2.203717472118959in,]{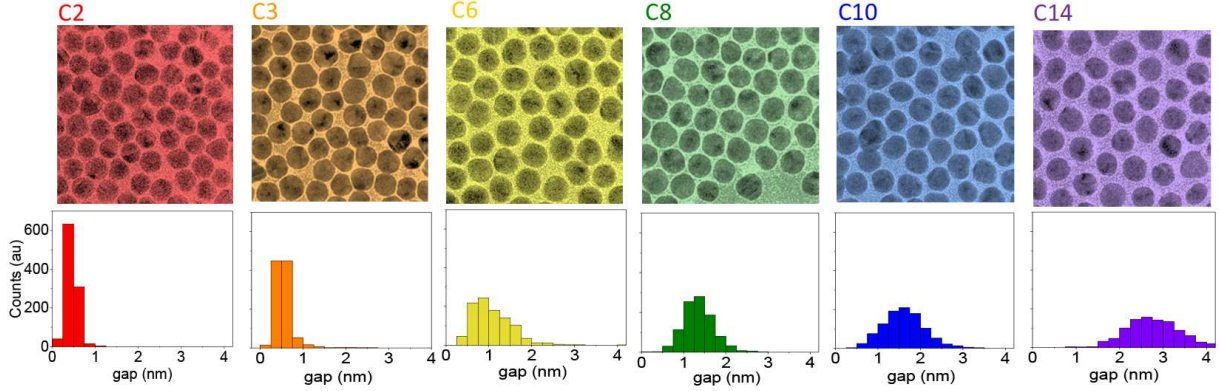}
}
\caption{ False-colored TEM images (top) and interparticle gaps distributions (bottom) for ligand mixtures; C2 (1-ethanethiol/1,2-ethanedithiol), C3 (1-propanethiol/1,3-propanedithiol), C6 (1-hexanethiol/1,6-hexanedithiol), C8 (1-octanethiol/1,8-octanedithiol), C10 (1-decanethiol/1,10-decanedithiol) and C14 (1-tetradecanethiol/1,14-tetradecanedithiol).}\label{2}\end{figure}

\begin{table}\centering \caption{
metasurface gaps
}
\setlength\fboxrule{0pt}\setlength\fboxsep{0pt}\fcolorbox[HTML]{000000}{FFFFFF}{
\begin{tabular}[c]{|c|c|c|}\hline
$ligand$ & $gap$ $\left (nm\right )$ & $\Delta gap$ $\left (nm\right )$ \\
\hline
$C2$ & $0.45$ & $ \pm 0.14$ \\
\hline
$C3$ & $0.55$ & $ \pm 0.22$ \\
\hline
$C6$ & $1.1$ & $ \pm 0.46$ \\
\hline
$C8$ & $1.4$ & $ \pm 0.37$ \\
\hline
$C10$ & $1.6$ & $ \pm 0.48$ \\
\hline
$C14$ & $2.8$ & $ \pm 0.69$ \\
\hline
\end{tabular}
}
\end{table}

The gap size depends on the number of carbon atoms in the alkane chains, $C$ (Figure 3a, black dots). The interparticle gap has been shown to increase at $0.12C$ $nm$ for hcp gold nanospheres coated with alkanethiols.\cite{RN1396,RN1395} If it is assumed the dithiols ligands set the gap size due to the additional $0.24$ $nm$ Au-S bond offset, relative to the monothiol ligand, and are interdigitated, then the gap size is expected to be of the form $0.12C +0.48$ $nm$ (Figure 3a, solid black line).  We find reasonable agreement between the theoretical prediction and the experimental data in Figure 3a, demonstrating the gap size is predominantly set by the length of the dithiol ligands.

As the interparticle gap decreases, the enhancement of the electric field in the gap dramatically increases leading to the absorbance peak wavelength red-shifting due to the increasing coupling strength between the individual nanosphere modes.\cite{RN309}  The normalized experimental absorbance peak spectra from each metasurfaces is shown in Figure 3b.   The absorbance peak shifts by $ \sim 175$ $nm$ covering yellow ($580$ $nm$ for C14) to near infrared red ($775$ $nm$ for C2) wavelengths.  The quality factor ($Q =peak$ $wavelength/FWHM$) of the absorbance peaks also remains large throughout.

Figure 3c relates the absorbance peak wavelength to the interparticle gap. The experimental peaks are represented by the black dots.  The uncertainty in the peak wavelength was determined by averaging measurements from 5 different spatial positions (beam area$ \sim 4$ $mm^{2}$) on $3$ separate metasurfaces for each ligand length.  The uncertainty in the gap sizes was determined from the TEM analysis (Table 1).

The onset for the breakdown of the electric field can be described using a nonlocal hydrodynamic model where the local electric field in the gap is attenuated through the introduction of a $\beta  -$parameter, which is proportional to the Fermi velocity of the electrons.  In the nonlocal model the electron current density, \textbf{J}, inside the metal induced by the applied electric field, \textbf{E}, oscillating at frequency, $\omega $, is given by,\cite{RN1415,RN1714,RN1715}\begin{equation}\beta ^{2} \nabla \left ( \nabla  \cdot \mathbf{J}\right ) +\left (\omega ^{2} +i\gamma \beta \right )\mathbf{J} =i\omega \omega _{p}^{2}\varepsilon _{0}\mathbf{E}
\end{equation}    where $\gamma $ is the dampening coefficient, $\omega _{p}$ is the plasma frequency of gold and $\varepsilon _{0}$ is the permittivity of free space (see Methods section for details).

\begin{figure}\centering 
\setlength\fboxrule{0.01in}\setlength\fboxsep{0.1in}\fcolorbox[HTML]{FFFFFF}{FFFFFF}{\includegraphics[ width=5.5in, height=2.2124925283921097in,]{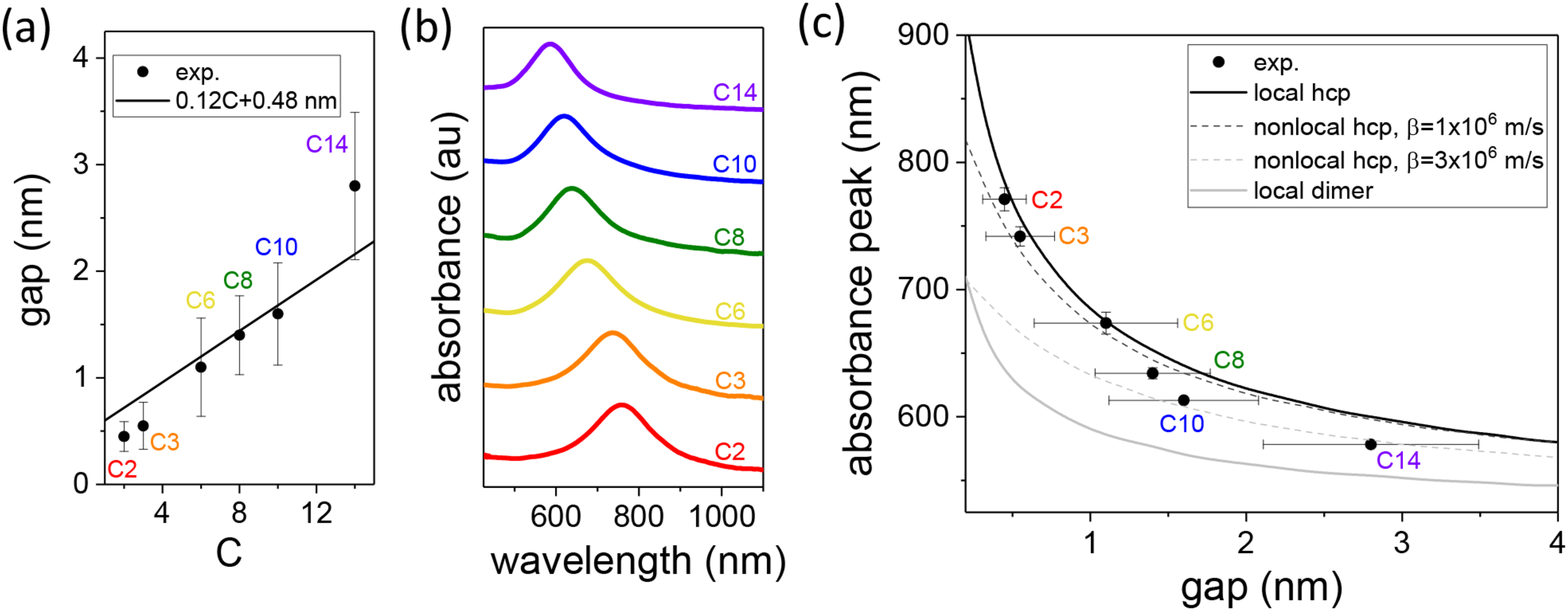}
}
\caption{ (a) Interparticle gap as a function of the number of carbon atoms, $C$, in the alkane chains. (b) Normalized experimental absorbance spectra of the metasurfaces for each ligand length. (c)  Experimental absorbance peak wavelength (black dots) of the metasurfaces as a function of interparticle gap.  Finite element simulations of the metasurface with hexagonally close packing of the nanospheres with a local response (thick solid black line), a nonlocal response (thin dotted black, $\beta  =1 \times 10^{6}$\ $m/s$; thin dotted grey, $\beta  =3 \times 10^{6}\ m/s$ lines), and local response from isolated dimers (thick solid grey line) as a function of interparticle gap.}\label{3}\end{figure}

To understand the experimental absorbance peak shift as the gap decreases three-dimensional finite-element simulations (COMSOL Multiphysics 5.3) of hcp metasurfaces with a classical (local) response and with the nonlocal response were carried out (see Methods section for details).  The thick solid black line in Figure 3c shows the model results of the hcp metasurfaces with a local response, setting an upper bound on the field enhancements and absorbance peak red-shift.  

Equation 1 was directly incorporated into the local hcp metasurface model, introducing the nonlocal effects.  The nonlocal response is shown in Figure 3c. Two values of $\beta $ were chosen; $\beta  =1 \times 10^{6}\ m/s$  (thin dotted black line) and $\beta  =1 \times 10^{6}\ m/s$  (thin dotted grey line). The nonlocal effects retrieved from the simulations for $\beta  =1 \times 10^{6}\ m/s$  start to become appreciable for gap sizes less than $ \sim 1$ $nm$.  As $\beta $ increases to $\beta  =3 \times 10^{6}\ m/s$ , the electric field inside the gap is further attenuated, leading to a reduction in the absorbance peak wavelength shift.  The experimental data is in good agreement with the nonlocal simulations for $\beta  =1 \times 10^{6}\ m/s$.\cite{RN1627}  On the contrary, when $\beta  =3 \times 10^{6}\ m/s$  the simulated data deviates significantly from the experimental results, especially for the subnanometer gap regime where nonlocal effects are expected to be triggered.

The local response of an isolated dimer was also modeled in Figure 3c (thick solid grey line).  The results show the significance of the six-fold near-neighbor coupling from the hcp symmetry, relative to the single near-neighbor in the dimer geometry, needed to red-shift the absorbance peak far enough to overlap with the experimental data.  The dimer evolution therefore places a lower bound on the absorbance peak shift.

For the larger gap sizes, $ >1$ $nm$ (C8-C14), the absorbance peak wavelength is blue-shifted relative to the local hcp metasurface model upper bound, yet shifted well above the local dimer lower bound.  Since the nonlocal effects are negligible for these gap ranges\cite{RN1627} it is assumed the reason is from non-ideal experimental hcp structures.  This assumption is supported by the data in Figure 2, showing the distribution of the gap sizes increases as the ligand length become longer. 

 As the gap decreases below $1$ $nm$ (C6-C2), the nonlocal models can theoretically deviate significantly from the local model depending on the value of $\beta $.  While the nonlocal effects are present, the maximum difference between the local model and the experimental data is only $14$ $nm$ for C2, less than a $2 \%$ difference.

The effective real part of the refractive index of a metasurface (C6) retrieved from the ellipsometry measurements is shown in Figure 4a.  The real part of the refractive index shows a minimum of $1.0$ at $555$ $nm$, a maximum of $5.0$ at $743$ $nm$ and remains larger than $3.5$ out to $1 ,550$ $nm$.    Both the magnitude and dispersion of the refractive index is large for these engineered metasurfaces, surpassing that of any known naturally occurring transparent material at visible and near infrared frequencies.\cite{RN1703,RN1719,RN1711}

\begin{figure}\centering 
\setlength\fboxrule{0.01in}\setlength\fboxsep{0.1in}\fcolorbox[HTML]{FFFFFF}{FFFFFF}{\includegraphics[ width=6.5in, height=2.1398377455166524in,]{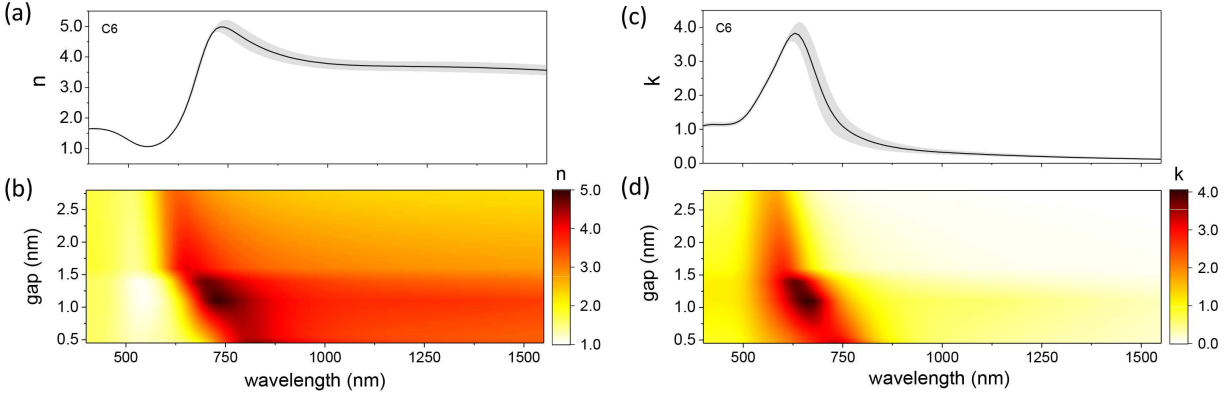}
}
\caption{Real (a) and imaginary (c) parts of the effective refractive index of a metasurface (C6) obtained from spectroscopic ellipsometry measurements (solid black line). The grey area is the uncertainty.  Real (b) and imaginary (d) parts of the effective refractive index of the metasurfaces as a function of ligand length.
}\label{4}\end{figure}

Figure 4b shows the evolution of $n$ as a function of gap.  As expected, the spectra red-shifts in wavelength as the length of the ligand decreases from C14 to C2.  The largest $\left (5.0\right )$ refractive index occurs for a gap of $1.1$ $nm$ (C6) and not at the smallest gap of $0.45$ $nm$ (C2) as expected if the response was purely local.  From Figure 2, C6 has a larger gap size distribution, relative to C3 and C2, implying a lower quality hcp structure, consequently the extreme enhancement for C6 is not due to a reduction in packing efficiency for gap sizes below $1$ $nm$.  The C6 gap length scale is consistent with the established local-to-nonlocal crossover regime, $ \sim 1$ $nm$;\cite{RN1415} therefore, a plausible explanation for the extreme enhancement for the C6 metasurface is that an upper bound has been reached for the electric field enhancement, and resulting refractive index, before nonlocal effects become noticeable, in agreement with Figure 3.

The effective imaginary part of the refractive index of a metasurface (C6) retrieved from the ellipsometry measurements is shown in Figure 4c.  The peak maxima occurs at $629$ $nm$ with a quality factor of $4.3$. The figure of merit, $n/k$, at the peak maxima peak is $0.7$.  However, the imaginary part of the refractive index drops close to zero approaching near infrared wavelengths, while the real part of the refractive index remains large. As a consequence, the figure of merit is $28.5$ at $1 ,550$ $nm$, forty times larger than at visible wavelengths, which may have useful telecommunication application consequences.\cite{RN1718}  Figure 5d shows the peak of the imaginary part of the refractive index red-shifting as the gap decreases, in agreement with absorbance data collected in Figure 3c.\cite{RN309}

Subtle differences in ligand composition are known to dramatically influence the electrical properties through the ligand at low frequencies, e.g. the electrical conductivity decreases by four orders of magnitude if di-thiol alkane ligands are replaced with mono-thiols.\cite{RN1704}   An unresolved issue is the effect ligands have at higher frequencies on the optical response.\cite{RN1627}

To determine how the terminal ends of the ligands influence the optical response of the metasurfaces, ligands with alkane or thiol end groups were exchanged. The ligand shells were composed solely of mono- (1-propanethiol (C3-mono)) or di-thiols (1,3-propanedithiol (C3-di)) and compared to the standard mixed ligand shells (1-propanethiol/1,3-propanedithiol (C3)).  Furthermore, to understand the role of ligand conjugation, the alkane ligand cores were replaced with a highly conjugated core; 1,4-benzenedithiol (BZT).

Figure 5a shows the experimental absorbance spectra for the C3, C3-di, C3-mono and BZT metasurfaces.  Unsurprisingly, the C3-di absorbance spectrum closely matches the C3 spectrum, since the di-thiol ligands set the gap size, with similar peak wavelengths and quality factors, see Table 2.

\begin{figure}\centering 
\setlength\fboxrule{0.01in}\setlength\fboxsep{0.1in}\fcolorbox[HTML]{FFFFFF}{FFFFFF}{\includegraphics[ width=5.5in, height=2.320224719101124in,]{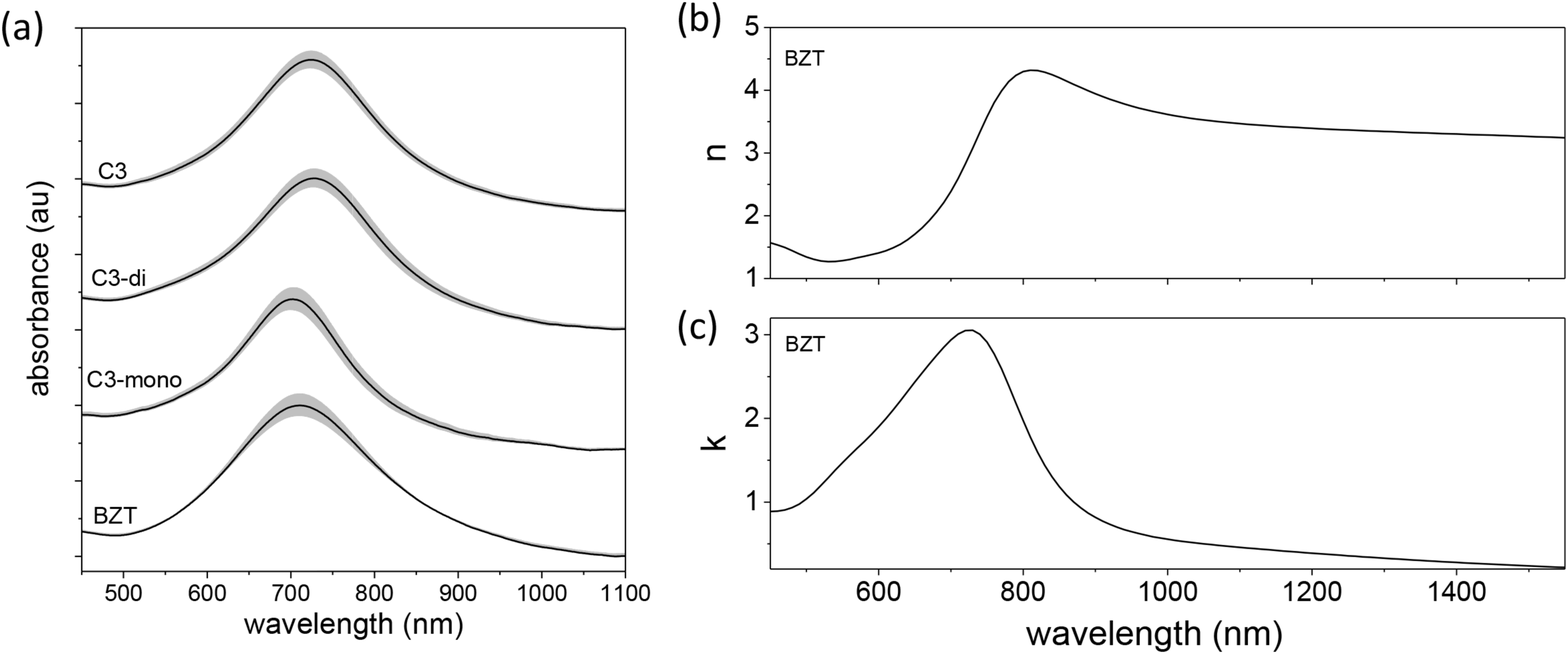}
}
\caption{(a) Normalized experimental absorbance spectra for the C3, C3-di, C3-mono and BZT metasurfaces.  The grey area is the uncertainty.    Effective real (b) and imaginary (c) refractive index of the BZT metasurface. 
}\label{4}\end{figure}
\begin{table}\centering \caption{
metasurface ligands
}
\setlength\fboxrule{0pt}\setlength\fboxsep{0pt}\fcolorbox[HTML]{000000}{FFFFFF}{
\begin{tabular}[c]{|c|c|c|c|}\hline
$ligand$ & $abs .$ $peak$ $\left (nm\right )$ & $FWHM$ $\left (nm\right )$ & $Q$ \\
\hline
$C3$ & $724$ & $191$ & $3.8$ \\
\hline
$C3 -di$ & $728$ & $191$ & $3.8$ \\
\hline
$C3 -mono$ & $702$ & $168$ & $4.1$ \\
\hline
$BZT$ & $710$ & $215$ & $3.3$ \\
\hline
\end{tabular}
}
\end{table}

For the C3-mono case in Figure 5a, the peak wavelength is slightly blue shifted by $3 \%$$(22\ nm)$ relative to the C3 peak, which is expected if tunneling is occurring.  However, the quality factor for C3-mono is larger $\left (4.1\right )$ compared to the C3 metasurface $\left (3.8\right )$, which is not expected if tunneling is occurring, since it would broaden the C3-mono peak and decreases the quality factor less than C3.\cite{RN1627,RN1714,RN1716,RN1017}  Moreover, it is well established that mono-thiols are significantly more insulating than di-thiols at low frequencies in molecular junctions.\cite{RN1704}  This relationship is expected to hold at elevated frequencies.\cite{RN1629}

Entropic forces are anticipated to dictate the packing of the C3-mono nanospheres during the metasurface assembly process, due to the absence of the additional thiol.  Therefore, it is reasonable the C3-mono ligands are not interdigitated, as in the C3 and C3-di cases, leading to the gap being approximately twice the ligand length.  The C3 gap was experimentally determined to be $0.5 \pm 0.22$ $nm$ (Table 1).  If it is assumed the C3-mono gap is twice the C3 gap or approximately $1$ $nm$, then the absorbance peak wavelength is in reasonable agreement to the local response model in Figure 3c. If the gap is assumed to be only one ligand length in size, or approximately $0.6$ $nm$, it would correspond closely to the $\beta  =3 \times 10^{6}\ m/s$ evolution, in disagreement with all the other experimental evidence.

Significantly effort was made to transfer the C3-mono metasurfaces to TEM grids to enable direct imaging of the gaps.  Due to a lack of an attractive di-thiol mechanism between the nanospheres during the self-assembly process, the C3-mono nanospheres packed very poorly on the TEM, leading to unreliable gap distribution measurements.  Without the availability of reliable TEM imagery direct conclusions cannot be drawn on the uniformity of the hexagonally close packing, yet if the packing were not uniform, the quality factor would be expected to be smaller than the C3 metasurfaces.  This indirect evidence implies the C3-mono metasurface are well packed with uniform gaps.  Based on all the evidence above, the C3-mono metasurfaces are unlikely to be transferring charge at optical frequencies.

The BZT metasurface absorbance spectrum is shown in Figure 5a.  The BZT absorbance peak wavelength is blue shifted by $2 \%$ $(14\ nm)$ relative to the C3 metasurface, but red shifted by  $\left (8\ nm\right )$compared to the C3-mono metasurface (Table 2), which falls within the experimental uncertainty window.  The quality factor for the BZT metasurface is $13 \%$ smaller than the C3 and $20 \%$ smaller than the C3-mono metasurfaces, due to an increased $FWHM$.  The increased $FWHM$ can be a consequence of non-uniform hcp packing, or it may also be a consequence of tunneling.\cite{RN1627,RN1714,RN1716,RN1017} As in the C3-mono case, reliable gap distribution measurements were not possible for the BZT metasurfaces due to inability to transfer the metasurfaces well onto the TEM grids.

To unravel the origins of the increased $FWHM$ for the BZT metasurface the effective real and imaginary refractive indices were determined from spectroscopic ellipsometry measurements and are shown in Figure 5b and 5c, respectively.  A previous experimental report demonstrated that if charge is transferred between nanoparticles interconnected with BZT, then a secondary near infrared absorbance peak emerged.\cite{RN1629}  Theoretical investigations for nanospsheres bridged with BZT also showed the primary absorbance peak blue shifts and broadens and a secondary near infrared absorbance peak can emerge, but is highly dependent on the conductivity of the molecular junction and geometry of the nanoparticles.\cite{RN1627}  From the imaginary part of the refractive index in Figure 5b we find no evidence of a secondary near infrared absorbance peak as far out as $1 ,550$ $nm$.  Based on the minimal absorbance peak blue shift and slight broadening, as well as the lack of an additional absorbance peak, these results suggest the BZT metasurfaces are also not likely to be transferring charge at optical frequencies and the increased $FWHM$ is a consequence of non-uniform nanosphere packing.

\textbf{CONCLUSIONS}

In summary, using large-area, self-assembled metasurfaces with tunable interparticle gaps approaching length scales comparable to the atomic lattice constant of gold, we experimentally showed that the optical response agrees well with a classical model. We demonstrated that the optical response of these metasurfaces are relatively inert to the degree of conjugation and terminal end groups of the ligands, finding no conclusive evidence of charge transfer at optical frequencies.  These results imply that a straightforward classical model is sufficient to describe the linear optical response of most pragmatic plasmonic systems.  Finally, we demonstrate a upper bound on the confinement of light using these metasurfaces, leading to optical properties well beyond that of any known naturally occurring transparent material at visible and near infrared wavelengths, with the effective real part of the refractive index as small as $1.0$ at $555$ $nm$ and as large as $5.0$ at $743\ $$nm$.

\textbf{METHODS}

\textit{Assembly. } Gold nanospheres were also commercially acquired from Ted Pella, Inc. To direct the assembly of the metasurfaces an aqueous suspension of gold nanospheres, at an optical density of OD$ \sim 1$ in $1$$\ cm$ (measured at the abs. peak) and stabilized in a sodium citrate solution, is mixed into a $20$ $ml$ scintillation vial containing a tetrahydrofuran solution $\left (1\ ml\right )$ with mono- and di-thiol alkane ligands in equal proportions $(10\ \mu l\ of\ ea .)$.  For the ethanethiol cases the ligand volumes were doubled.  Upon thoroughly mixing, glass substrates (cleaned in potassium hydroxide dissolved in methanol at a concentration of $5 \%$(w/v)) were then placed into the vial and wetted with the reaction solution. The ligands begin to bind to the surface of the nanospheres causing the ligand coated nanospheres to phase separate from the water and tetrahydrofuran suspension to the air-fluid interface.  Surface tension gradients, arising from different vapor pressures between the water and tetrahydrofuran, carry the functionalized nanoparticles onto a pre-wetted glass substrate creating macroscopic monolayer metasurfaces. The process was repeated for each ligand type and the metasurfaces were dried overnight. For the metasurfaces assembly process with solely mono- or di-thiols, $20$$\ \mu l$ of each were used and $5$ $mg$ for the benzenedithiol case.\cite{RN590,RN1711}

\textit{Characterization.} An unpolarized white light source (HL-2000) was used to probe the metasurfaces and the light was collected with optical fiber (OceanOptics QP400-1-VIS-NIR) coupled into a spectrophotometer (OceanOptics QEPro).  Transmission electron microscopy (TEM) images were obtained using a JEOL JEM-2200FS field emission electron microscope with $200\ $$kV$ accelerating voltage.  TEM samples were prepared by placing a $ \sim 10$$\mu l$ drop of the sample suspension onto carbon coated copper grid, wicked off with lens paper, and allowed to dry overnight.  The effective real, $n$, and imaginary, $k$, parts of the refractive index of the metasurfaces were determined as a function of gap from ellipsometry measurements (J.A. Woollam Co., V-VASE).  The reflection measurements were measured off the air-nanosphere interface at an angle of $60^{ \circ }$ with wavelengths ranging $400 -1 ,550\ $$nm$ and scotch tape was placed on the backside (glass-air) surface to scatter any secondary reflections.  By comparing the measured spectroscopic ellipsometry angles $\Psi $ and $\Delta $ (averaged over $5$ different spatial positions for $3$ metasurfaces per gap size with a beam area of $7\ $$mm^{2}$) to a $3$ layer model, composed of the glass substrate, gold nanospheres and an ambient layer the optical response for each metasurface was determined.  The modeled layer was assumed to be homogenous since it is deeply sub-wavelength and was set to a thickness equal to the diameter of the gold nanospheres, $13.3$$\ nm$, plus twice the ligand length.  Gaussian oscillators were selected to model the dielectric function as they ensure a Kramers-Kronig consistent line shape.  A total of $6$ Gaussian oscillators were used to fit, along with $1$ offset and $2$ poles located outside of the collected data.

\textit{Simulations.}  The local and nonlocal numerical full-wave simulations were carried out using the commercial finite-element simulation software COMSOL Multiphysics. The symmetry and periodicity of the proposed metasurface was used to reduce the three-dimensional simulation domain. The metasurface was illuminated with linear polarized waves and the reflectance $\left (R\right )$ and transmittance $\left (T\right )$ spectra were computed. The calculated absorbance spectrum was given by $A =1 -R -T$. In addition, we assumed that the periodically arranged gold nanospheres were coated with alkanethiol ligand with refractive index $1.44$. The metasurface is placed on top of a semi-infinite glass substrate with refractive index $1.53$, similar to the experimental set-up.  The nonlocal electromagnetic response of metals (gold in this case) can be described by the hydrodynamic model.\cite{RN1415,RN1720} In this case, the nonlocal free electron current density $\mathbf{J}$ inside the metal induced by the applied electric field, $\mathbf{E}$, oscillating at frequency $\omega $, is given in frequency domain by the equation: $\beta ^{2} \nabla \left ( \nabla  \cdot \mathbf{J}\right ) +\left (\omega ^{2} +i\gamma \beta \right )\mathbf{J} =i\omega \omega _{p}^{2}\varepsilon _{0}\mathbf{E}$, where the dampening coefficient $\gamma $ and the plasma frequency $\omega _{p}$ of the dispersive permittivity of gold were obtained from experimental data\cite{RN1268} and the $\beta $ parameter is proportional to the Fermi velocity of electrons in gold.\cite{RN1415} The nonlocal current density $\mathbf{J}$ is computed by using the weak form partial differential equation module of COMSOL. Then, the calculated current density $\mathbf{J}$ is introduced as a weak contribution to the frequency domain electromagnetic solver of COMSOL. More specifically, this weak contribution is added as an additional polarization term in the electromagnetic wave equation: $ \nabla  \times  \nabla  \times E -k_{0}^{2}E =i\omega \mu _{0}J$ , which is solved by the frequency domain electromagnetic solver of COMSOL. Finally, the continuity of the normal component of the electric displacement field between gold and the surrounding dielectric (ligand) needs to be manually introduced at the weak form module of COMSOL as an additional Dirichlet boundary condition. This is due to the presence of spatial derivatives in the nonlocal free electron current density equation.          
            
\textbf{Acknowledgement}

We (N.C., J.N., S.T., A.P., B.R., J.F.) thank the Office of Naval Research for support.   D.D. was a 2017 summer student under the Naval Research Enterprise Internship Program (NREIP) and is an undergraduate student at the University of Pittsburgh.  J.H. and C. A. were 2017 fellows in the Office of Naval Research Summer Faculty Research Program and are faculty at the University of Arkansas and University of Nebraska-Lincoln, respectively.  R.N. acknowledges the National Research Council for a postdoctoral fellowship.

\ \bibliographystyle{apsrev4-1}
\bibliography{}
\end{document}